\documentclass[%
 reprint,
 amsmath,amssymb,
 aps,
prb,
]{revtex4-1}

\usepackage{setspace} 
\usepackage[sc]{mathpazo} 
\usepackage[T1]{fontenc} 
\usepackage[utf8x]{inputenc} 
\usepackage{numprint} 
\usepackage{siunitx} 
\usepackage{graphicx} 
\usepackage[a4paper]{geometry} 
\usepackage{cleveref} 
\usepackage{fancyhdr} 
\usepackage{epigraph}
\usepackage{ctable} 
\usepackage{xcolor} 
\usepackage[english]{babel} 
\usepackage[multiple,stable,bottom]{footmisc} 
\usepackage{lipsum} 
\usepackage{microtype} 
\usepackage{geometry} 
\usepackage{amsmath} 
\usepackage{amssymb} 

\usepackage{longtable}
\usepackage{bigstrut}
\usepackage{enumerate}
\usepackage{todonotes}

\usepackage{float} 

\newcommand*{\citen}[1]{%
  \begingroup
    \romannumeral-`\x 
    \setcitestyle{numbers}%
    \cite{#1}%
  \endgroup   
}

\usepackage{dcolumn}
\usepackage{bm}

\begin{document}

\preprint{APS/123-QED}

\title{Scanning Kerr microscopy study of current induced switching in Ta/CoFeB/MgO films with perpendicular magnetic anisotropy}

\author{C. J. Durrant}
\author{R. J. Hicken}%
\affiliation{Department of Physics and Astronomy, University of Exeter, Exeter, EX4 4QL UK}

\author{Qiang Hao}
\author{Gang Xiao}
\affiliation{Department of Physics, Brown University, Providence, RI 02912 USA}

\date{\today}

\begin{abstract}
Ta/CoFeB/MgO trilayers with perpendicular magnetic anisotropy are expected to play a key role in the next generation of current and electric field switched memory and logic devices. In this study we combine scanning Kerr microscopy with electrical transport measurements to gain insight into the underlying mechanisms of current induced switching within such devices. We find switching to be a stochastic, domain wall driven process, the speed of which is strongly dependent on the switching current. Kerr imaging shows domain nucleation at one edge of the device which modelling reveals is likely assisted by the out-of-plane component of the Oersted field. Further domain growth, leading to magnetisation reversal, may still be dominated by spin torques.
\end{abstract}

\pacs{Valid PACS appear here}
\maketitle


\section{\label{sec:level1}Introduction}
The ever growing demand for higher density storage and faster processing speeds has led to great advances in the field of data storage in the last few decades\cite{Bandic2008}. The discovery of tunnelling magnetoresistance (TMR) has made the magnetic tunnel junction (MTJ) one of the leading candidates for next generation high density magnetic random access memory (MRAM).

Traditional MTJ memory requires an external magnetic field, generated by a current, to induce switching during the read/write procedure. Recently there has been great interest in switching the memory bit directly by using spin transfer torques (STT) generated by charge currents. Conventional STT switching requires current to be injected directly through a tunnel barrier but recent studies have utilised the torques generated by spin Hall and Rashba effects to switch elements with an in plane current. To date these phenomena have been investigated mostly in layered structures containing HM/FM/Ox sub-units in which the ferromagnetic (FM) layer is sandwiched between a heavy metal (HM) and an oxide (Ox). Spin-transfer torque magnetic random-access memory (STT-MRAM), has been explored as a viable option to replace dynamic random access memory (DRAM) and may provide performance comparable to DRAM main memory with an average 60\% reduction in main memory energy\cite{Kultursay2013}. With main memory energy now accounting for as much as 30\% of overall system power \cite{Carter2010, Hoelzle2009} STT-MRAM has the potential to significantly reduce the operational cost of computing systems.

It has been demonstrated that the Ta/CoFeB/MgO sub-unit could fulfil the three criteria required for high performance MTJs for STT-MRAM, namely high tunnelling magnetoresistance (TMR), low switching current, and high thermal stability for small device dimensions. With TMR ratios of 120\% in Ta/MgO/CoFeB/MgO/CoFeB/Ta\cite{Ikeda2010} stacks and 162\% in similar stacks with Mo layers\cite{Almasi2015}, along with switching currents of $3-6 \times 10^6$ A/cm$^2$ for similar devices\cite{Qiu2014,Zhang2014}, and good thermal stability for devices with $40$ nm diameter\cite{Ikeda2010}, Ta/CoFeB/MgO is indeed a viable candidate for next generation STT-MRAM.
 
A deeper understanding of the underlying switching mechanisms  is critical for effective device fabrication and has recently been the subject of much debate. Kim \textit{et al}\cite{Kim2013} report a strong HM and FM thickness dependence of both the Rashba spin orbit field and the spin Hall torques. Torrejon \textit{et al}\cite{Torrejon2014} also report a strong thickness dependence of their relative contributions to the switching, with spin Hall torque dominating for thicker HM underlayers. The relative contributions made by these effects are also influenced by capping layer thickness\cite{Qiu2013} and temperature\cite{Qiu2014}. In this study we combine scanning Kerr imaging with electrical transport measurements to gain further insight into the switching process.

\section{\label{sec:level1}Experimental}
\subsection{\label{sec:level2}Fabrication Details}
We focus only on CoFeB with the composition Co$_{40}$Fe$_{40}$B$_{20}$ in which perpendicular magnetic anisotropy (PMA)
was been achieved through thermal annealing. Stacks of Ta(4 nm)/CoFeB(1 nm)/MgO(1.6 nm)/Ta(1 nm) were sequentially deposited on thermally oxidised Si wafers in a high vacuum magnetron sputtering system. Further details of this process can be found in ref. \citen{Hao2015}. The 1 nm Ta layer was used as a capping layer to prevent atmospheric oxidisation.  

The stacks were patterned into Hall bars (20 x 140 $\mu$m\textsuperscript{2}) (Fig.\ref{Figure-CoFeB-Images Coercive Fields}.(d)) using photolithography. After stack deposition a second round of photolithography and a Au deposition was used to define contact pads. Au contacts were not critical for switching but allow wire-bonding to facilitate electrical measurements.  

The sample was then annealed in vacuum (1 x 10\textsuperscript{-6} Torr) at $220\,^{\circ}\mathrm{C}$ for 1 hour with 2 hours of ramping up and 6 hours of natural cooling under a magnetic field of 0.45 T perpendicular to the plane of the sample. It was shown by several groups\cite{Hao2015, Worledge2011, Avci2014} that the $220\,^{\circ}\mathrm{C}$ annealing temperature is critical for the formation of robust PMA. As-prepared samples do not support PMA as they have sharp, but disordered, interfaces while at high annealing temperatures (>$220\,^{\circ}\mathrm{C}$) diffusion within the interfacial region is detrimental to the PMA.

\subsection{\label{sec:level2}Measurement Details}

For magnetotransport measurements wire-bonding was used to connect the Hall bar to two 50 $\Omega$ coplanar waveguides (CPWs). A current was applied along the direction parallel to the long edge of the Hall bar ($\mathbf{\hat{z}}$) and the Hall resistance was measured across the contacts perpendicular to the current ($\mathbf{\hat{x}}$) as shown in Fig.\ref{Figure-CoFeB-Images Coercive Fields}.(d). In this way the magnetization between the contact pads was probed via the anomalous Hall effect (AHE). The Hall bar and CPWs were placed in a scanning Kerr microscope so that the magnetisation within a sub-micron region could be probed via the magneto-optical Kerr effect (MOKE), simultaneously with the Hall resistance measurements. The MOKE measurement was performed by focusing the beam from a 633 nm He-Ne laser onto the sample surface with a 40X objective lens, and recording the optical rotation of the back-reflected beam using a simple optical bridge detector. 
In each experiment a large in plane field was used to set the initial perpendicular magnetisation direction. The direction of the magnetisation was reproducible due to a slight ($< 1\deg$) tilt of the applied field relative to the plane of the Hall bar. The field was then reduced to remanence ($\approx7$ Oe) along the direction of current flow, as shown in Fig.\ref{Figure-CoFeB-Images Coercive Fields}.(c). All measurements were performed at room temperature.

\section{\label{sec:level1}Results and Discussion}

\begin{figure}[ht!]
	\centering
	\includegraphics[scale=0.8]{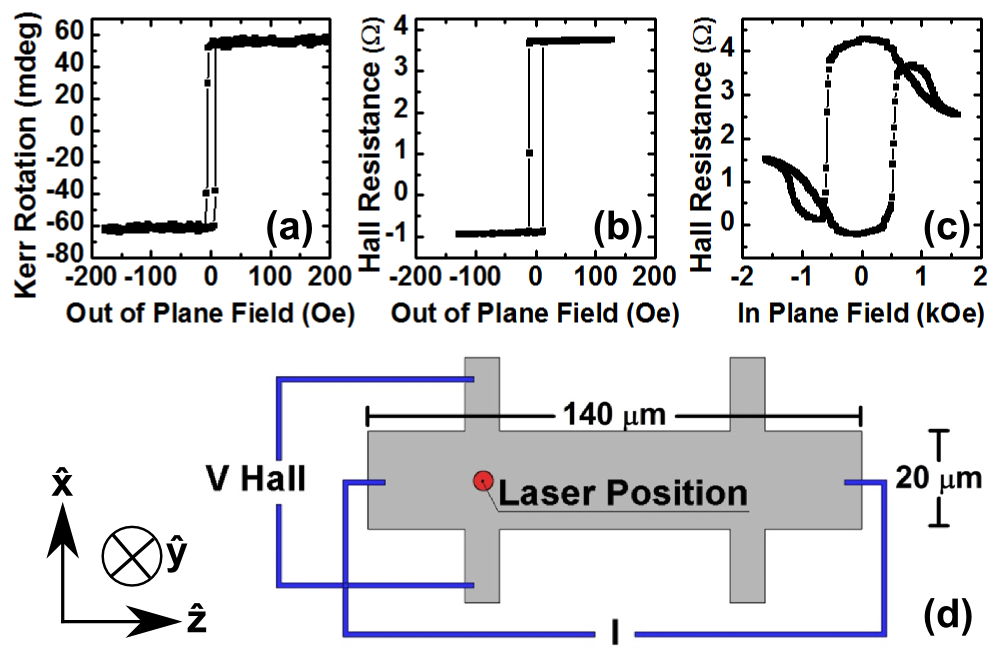}
	\caption{Kerr Rotation (a) measured mid-way between the contact pads as shown in (d), and Hall resistance (b) between the contact pads during sweeping of an out-of-plane ($\mathbf{\hat{y}}$) magnetic field. (c) shows the Hall resistance measured during sweeping of an in-plane magnetic field applied along the direction of the current ($\mathbf{\hat{z}}$).}
	\label{Figure-CoFeB-Images Coercive Fields}
	\end{figure}

Using an out-of-plane ($\mathbf{\hat{y}}$) field the saturation Kerr rotation (Fig.\ref{Figure-CoFeB-Images Coercive Fields}.(a)) and Hall resistance (Fig.\ref{Figure-CoFeB-Images Coercive Fields}.(b)) were found for each of the bistable out-of-plane magnetisation \textbf{M} states. These results were used to confirm full magnetization reversal had occurred in subsequent current induced switching experiments. The devices showed a particularly low ($\approx$ 10 Oe) perpendicular coercive field. The Hall resistance for the in-plane field used to set the initial \textbf{M} state is shown in Fig.\ref{Figure-CoFeB-Images Coercive Fields}.(c).

\subsection{\label{sec:level2}Time resolved - time-dependent current}
\begin{figure}[ht]
	\centering
	\includegraphics[scale=0.8]{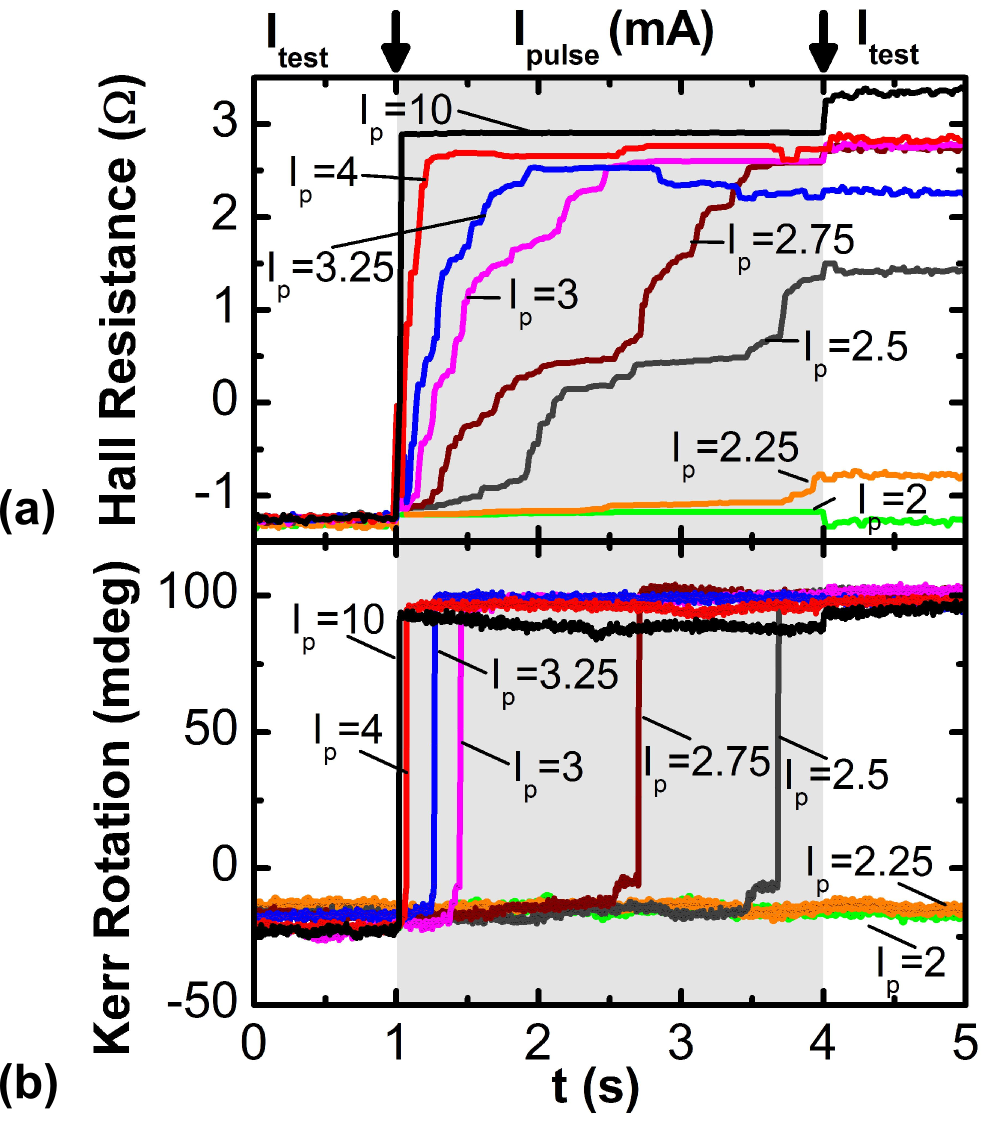}
	\caption{Hall resistance (a) between the contact pads shown in Fig.\ref{Figure-CoFeB-Images Coercive Fields}.(d) as the device undergoes magnetization switching induced by current pulses of 3s duration with amplitude $I_p$ = 2.00-10.00 mA, triggered at $t$ = 1 s (darker background on graph). Hall resistance is measured using a current of 0.1 mA both before ($t$ = 0.0-1.0 s) and after ($t$ = 4.0-5.0 s) application of the pulse. (b) shows the simultaneous measurements of the Kerr rotation within a sub micron region mid-way between the Hall contacts, also shown in Fig.\ref{Figure-CoFeB-Images Coercive Fields}.(d).}
	\label{Figure-CoFeB-Varying current switching}
	\end{figure}

In the magnetotransport measurements shown in Fig.\ref{Figure-CoFeB-Varying current switching}.(a) the initial magnetisation state (\textbf{M}$^-$) was measured with a small D.C. current $I_{test} = 0.10$ mA and was found to correspond to a Hall resistance of $-1.25$ $\Omega$. The magnetization was switched towards the opposite bistable state (\textbf{M}$^+$), which corresponds to a Hall resistance of $3.25$ $\Omega$, by current pulses $I_p = 2.00-10.00$ mA with a duration of $3.0$ s. $I_{test}$ was then applied once more to measure the final magnetisation state. Throughout this process the Kerr rotation (Fig.\ref{Figure-CoFeB-Varying current switching}.(b)) was continuously monitored at a position mid-way between the Hall contacts shown in Fig.\ref{Figure-CoFeB-Images Coercive Fields}.(d). 

For $I_p>6.00$ mA, switching occurred on a timescale < 1 $\mu$s, faster than the resolution of the measurement technique. By reducing the pulsed current to the range $I_p = 2.25-5.00$ mA switching occurs considerably more slowly allowing the change in Hall resistance to be easily observed within the $3$ s pulse duration. Due to the strong PMA, an intermediate Hall resistance value corresponds to a domain state in which the magnetisation points either in to or out of  the plane. The intermediate Hall resistance is effectively equivalent to a line integral of all \textbf{M} states between the contact pads. This is confirmed by the Kerr rotation shown in Fig.\ref{Figure-CoFeB-Varying current switching}.(b) where the local \textbf{M} switches instantaneously between the two states. Switching in Fig.\ref{Figure-CoFeB-Varying current switching}.(b) coincides with large steps in the Hall resistance, which are most clear for currents in the range $I_p = 2.50-4.00$ mA. This implies that a large area (length equal to a few $\mu$m) switches simultaneously. Whether the switching is due to a single domain or a collection of smaller domains is difficult to determine as for $I_p = 2.50$ and $2.75$ mA a small step before full switching can be seen in the Kerr rotation (Fig.\ref{Figure-CoFeB-Varying current switching}.(b)). This must be due to either the full switching of a small domain, less than the size of the laser spot, or else the laser has been positioned on a domain wall at the upper edge of a larger domain. For the highest currents (seen for $I_p=10.00$ mA in Fig.\ref{Figure-CoFeB-Varying current switching}.(a)), the final Hall resistance is slightly larger than that observed for lower currents. This is likely due to the magnetization of the contact pad regions being more strongly pinned and requiring higher currents to switch than the body of the device. For higher currents (again seen clearly for $I_p = 10.00$ mA) the Hall resistance shows full switching occurs within the first $\mu$s but the final Hall resistance measured with $I_{test}$ is larger than the Hall resistance measured during the pulse. This behaviour may be due to the perpendicular component of the Oersted field (in the $\mathbf{\hat{y}}$ direction) opposing reversal on one edge of the device during $I_p$. This field will be significantly lower during $I_{test}$, when full reversal finally occurs. The role of the Oersted field will be discussed further in the final section of this paper. For the lowest currents $<2.25$ mA no switching occurs, which allows us to calculate the critical switching current $I_c$ for this device. The switching process is stochastic in nature as shown in Fig.\ref{Figure-CoFeB-Same current stochastic} so an average of repeated measurements gave $I_c \approx 2.6$ mA. Many comparable studies attribute switching purely to the giant spin Hall effect and so calculate critical current densities from only the current in the Ta layer. Based on typical resistivities of $\rho_{CoFeB} = 100$ $\mu \Omega $cm and $\rho_{Ta} = 200 \mu \Omega $cm \cite{Hao2015}, $\frac{2}{3}$ of the current flows in the Ta layer. This gives $I_c = 1.73$ mA and a critical current density of $2.17 \times 10^6$ A/cm$^2$. In stochastic systems critical parameter values depend upon both temperature and the characteristic measurement time however it is still worthwhile to compare this critical current density to similar studies\cite{Qiu2014,Zhang2014} which report critical densities of $3-6 \times 10^6$ A/cm$^2$. The critical switching current for this PMA device is comparable to those reported for the better established in plane MTJs\cite{Ikeda2007} and giant magnetoresistance (GMR) spin-valve sensors\cite{Lacour2004}. 
Our experiment not only demonstrates the strong dependence of the switching speed on $I_p$ but also highlights the need for careful consideration of current sweep rates when extracting $I_c$. Close to $I_c$ partial switching can occur, which may artificially lower $I_c$ values extracted from swept current experiments in which \textbf{M} is not reset between incremental increased of the current values.

\subsection{\label{sec:level}Time resolved - constant current}

\begin{figure}[ht]
	\centering
	\includegraphics[scale=0.9]{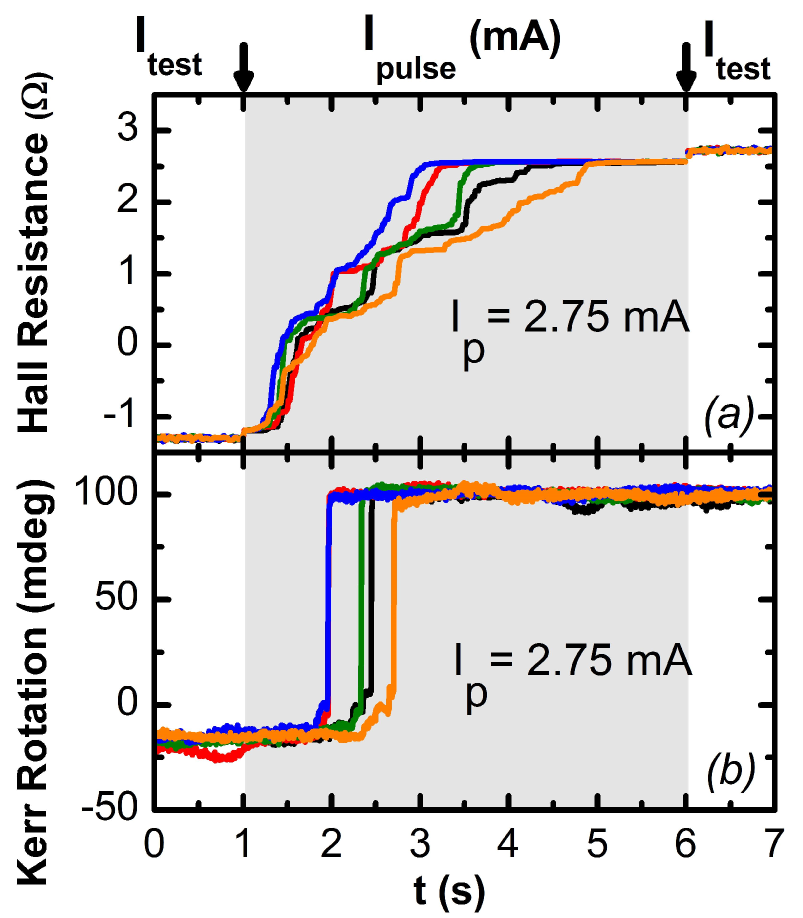}
	\caption{Hall resistance (a) between the contact pads shown in Fig.\ref{Figure-CoFeB-Images Coercive Fields}.(d) as the device undergoes $5$ switching events under identical conditions induced by $5.0$ s duration current pulses, with amplitude $I_p$ = $2.75$ mA, triggered at $t$ = $1.0$ s (darker background on graph). The Hall resistance is measured before ($t$ = $0.0-1.0$ s) and after ($t$ = $6.0-7.0$ s) the pulse using a current of $0.1$ mA. (b) shows the simultaneous measurement of the Kerr rotation within a sub micron region mid-way between the Hall contacts, also shown Fig.\ref{Figure-CoFeB-Images Coercive Fields}.(d).}
	\label{Figure-CoFeB-Same current stochastic}
	\end{figure}

During the experiments in which $I_p$ was varied, the Hall resistance trace and the time of switching observed in the Kerr signal were not always the same for repeated switching measurements made with identical $I_p$. Fig.\ref{Figure-CoFeB-Same current stochastic} highlights the stochastic nature of this process showing $5$ switching processes under nominally identical conditions for $I_p = 2.75$ mA. The parameters for this experiment were identical to those discussed for Fig.\ref{Figure-CoFeB-Varying current switching} but the pulse length was increased to $5.0$ s so that full switching could be obtained within the duration of the pulse at a lower $I_p$ value.

In the transport measurements shown in Fig.\ref{Figure-CoFeB-Same current stochastic}.(a) the Hall resistance is similar in each event for the first $1.0$ s of the pulse. The Hall resistance then diverges between $2.0-4.5$ s before reaching the same saturation state in the final $4.5-6.0$ s. The divergence coincides with a large step in the Hall resistance which occurs at the same time as the large change in Kerr rotation seen in Fig.\ref{Figure-CoFeB-Same current stochastic}.(b). As the Kerr rotation probes only the center of the device (Fig.\ref{Figure-CoFeB-Images Coercive Fields}.(d)), it can be inferred that for the first $1.0$ s of the pulse, domains at the edges of the Hall bar switch more easily than the magnetisation at its center, and that switching follows a similar 'path' in each event (reasons for this are to be discussed and shown in Fig.\ref{Figure-CoFeB-Images Modeling Oersted Fields}).  The switching of the central region, observed as a large step in Hall resistance and change in Kerr rotation, occurs later in the pulse and appears more random than the switching of the edge of the device.

\subsection{\label{sec:level} Static imaging during current induced switching}

\begin{figure}[ht!]
	\centering
	\includegraphics[scale=0.9]{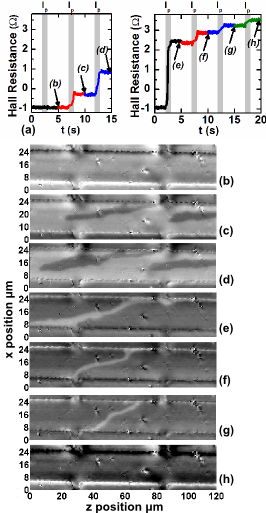}
	\caption{(a) Hall resistance values between the contact pads shown in Fig.\ref{Figure-CoFeB-Images Coercive Fields}.(d)  for the  magnetic domain state images shown in (b) to (h). Darker background in (a) indicates pulse on, lighter background indicates test current on.}
	\label{Figure-CoFeB-Images of cofeb switching}
	\end{figure}

For further insight into the domain configuration during switching, scanning MOKE images were acquired for a series of stable (on the timescale of the imaging experiment) domain configurations during the switching process. Fig.\ref{Figure-CoFeB-Images of cofeb switching}.(a) shows the Hall resistance values at which the images (b) to (h) were taken. It was not feasible to image a single switching event as the 'mid' states (with \textbf{M} close to 50/50 in and out of plane) were unstable over the imaging time (several hours). Images were instead taken on either side of this unstable middle region for the two separate partial switching events shown in Fig.\ref{Figure-CoFeB-Images of cofeb switching}.(a).
	
Fig.\ref{Figure-CoFeB-Images of cofeb switching}.(b) shows \textbf{M}$^-$ corresponding to a Hall resistance of $-1$ $\Omega$. As $I_p$ is applied we observe the formation (c) and growth (d) of large domains of \textbf{M}$^+$. These domains appear to grow from the edge of the Hall bar and remain pinned at more than one site (e.g. z $=60$ $\mu$m x $=24$ $\mu$m). For the second set of stable states the majority of \textbf{M} lies in the \textbf{M}$^+$ direction. The large reversed domain seen in (e) is in a similar position to the domain seen in (c) and (d) and also appears pinned at the same site. This domain then shrinks in (f) and (g) whilst remaining pinned until a full reversal to \textbf{M}$^+$ is observed in (h).

Reverse domains appear to form first at the edge of the device,  as observed in (b) to (d) where the domain forms at the edge (x $=24$ $\mu$m), and then grows towards the center. When reversal is close to completion the largest portion of the original domain state appears to be located on the opposite edge of the device, as seen clearly in (e) around x = $4$ $\mu$m. This may again be due to the presence of the Oersted field, with the $\mathbf{\hat{y}}$ component of this field ($B_y$) having peak magnitude but opposite polarity at the long edges of the device.

\subsection{\label{sec:level}Modelling the Oersted field}

Many studies of similar structures do not discuss the presence of Oersted fields. Some studies\cite{Liu2012b, Yu2014} do calculate the in-plane component of the Oersted field ($B_x$) and conclude that it does not have any significant effect upon the switching since $B_x$ is about $1$ order of magnitude smaller than the effective fields generated by spin-torques (ref \citen{Liu2012b} calculates 0.3 Oe/mA for a 20 $\mu$m wide bar) and this field often acts to oppose the spin-torques\cite{Liu2012a}.

\begin{figure}[ht!]
	\centering
	\includegraphics[scale=0.8]{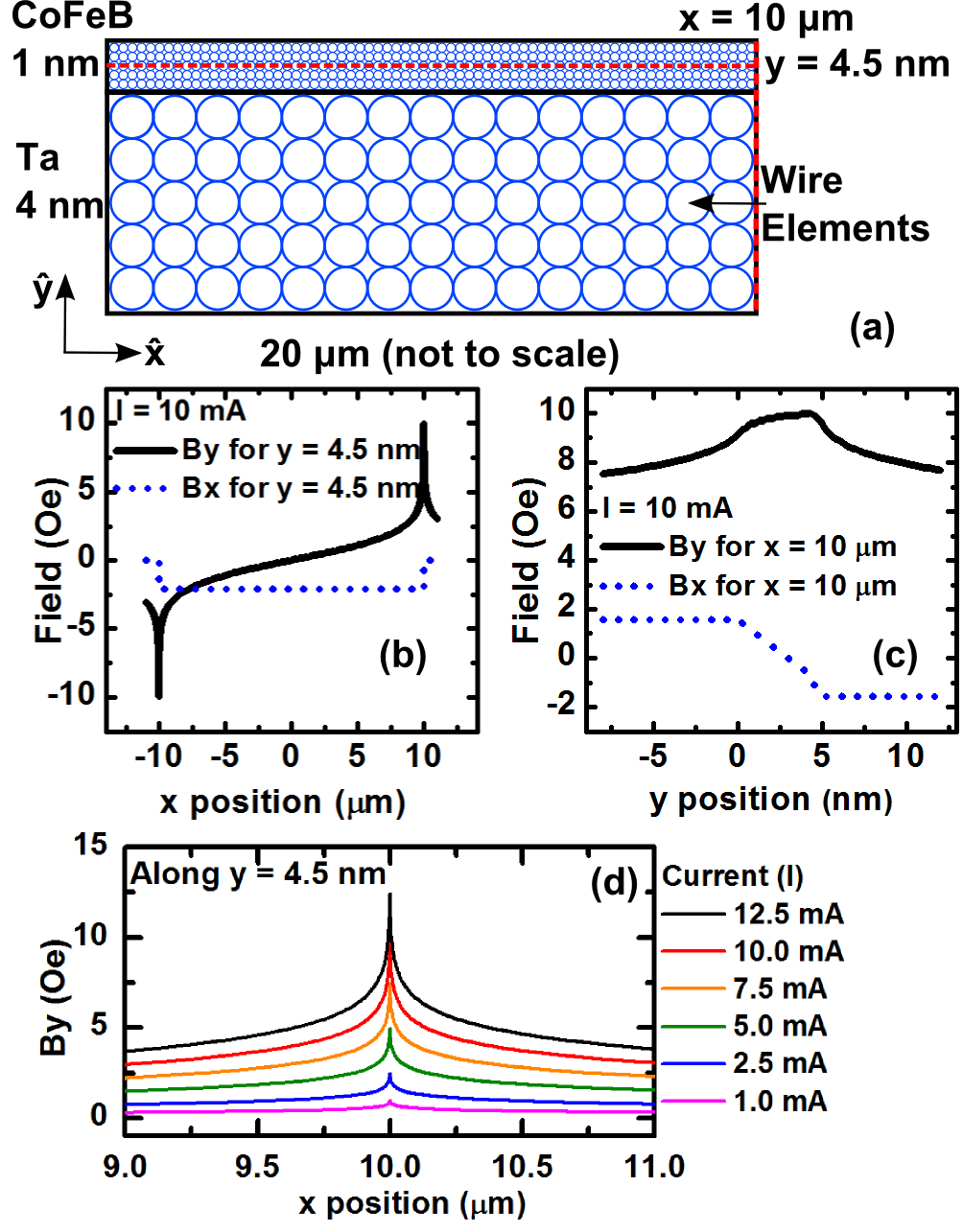}
	\caption{Geometry (a) for calculation of Oersted fields in Ta/CoFeB layers modelled as an array of wire elements. (b) shows how the in ($B_x$) and out-of-plane ($B_y$) components of the Oersted field vary across the width of the Hall bar along a line through the center of the CoFeB layer. (c) shows the variation of these fields across the thickness of the layers at the edge of the Hall bar where the out of plane field has maximum value. (d) shows how the out of plane field varies with current along a line through the center of the CoFeB layer close to the edge of the Hall bar.}
	\label{Figure-CoFeB-Images Modeling Oersted Fields}
	\end{figure}

  We demonstrate in the model shown in Fig.\ref{Figure-CoFeB-Images Modeling Oersted Fields} that whilst $B_x$ is indeed small $B_y$ can become comparable to the 10 Oe coercive field shown in Fig.\ref{Figure-CoFeB-Images Coercive Fields} and so may have a significant effect on the switching process, at least in the absence of an external field applied to the structure.

The Hall bar cross section ($20$ $\mu$m x $4$ nm Ta and $1$ nm CoFeB) was modelled by filling it with wires of radius $r$ as shown in Fig.\ref{Figure-CoFeB-Images Modeling Oersted Fields}.(a). The space in and around the bar was broken into a grid and the field calculated at each point from the Biot-Savart law	

%

A uniform current density was assumed in each layer. The current density in each layer was calculated assuming $\frac{2}{3} I$ in the Ta layer and $\frac{1}{3} I$ in the CoFeB (as discussed in the previous section), and  the current in an individual wire was adjusted to take account of the packing fraction. The current  was assumed to flow perpendicular to the cross sectional plane, shown in Fig.\ref{Figure-CoFeB-Images Modeling Oersted Fields}.(a) in the $\mathbf{\hat{z}}$ direction. Each layer was assumed to have a thickness equivalent to 5 wire diameters ($r_{CoFeB} = 1/10$ nm and $r_{Ta} = 4/10$ nm) in order to optimise computation time. Further increase in the number of wires contained within the layer made negligible difference to the calculated field. 

In  Fig.\ref{Figure-CoFeB-Images Modeling Oersted Fields} the $\mathbf{\hat{x}}$ direction has again been defined along the width of the Hall bar ($20$ $\mu$m) while $\mathbf{\hat{y}}$ lies perpendicular to the plane. Fig.\ref{Figure-CoFeB-Images Modeling Oersted Fields}.(b) shows $B_y$ and $B_x$ along a line through the center of the CoFeB layer at y $= 4.5$ nm for $I_p = 10$ mA. The field profile is as expected for a current carrying strip\cite{Keatley2008} and $B_x$ is in good agreement with fields calculated in ref.\citen{Liu2012b}. We observe a sharp peak in $B_y$ with height of about $\pm 10$ Oe close to the edges of the bar. These fields are comparable to the $10$ Oe out of plane coercive fields observed for this device in Fig.\ref{Figure-CoFeB-Images Coercive Fields}. The magnitude of the Oersted field varies little through the thickness of the CoFeB layer as shown at the field peak around (x $= 10$ $\mu$m), in Fig.\ref{Figure-CoFeB-Images Modeling Oersted Fields}.(c). These results in conjunction with Fig.\ref{Figure-CoFeB-Images of cofeb switching} may explain the domain behaviour at the edges of the device. On one edge $B_y$ aids the reversal of domains whilst on the opposite edge it opposes it. This effect may be lessened as $I_p$ and hence $B_y$ are reduced as shown in Fig.\ref{Figure-CoFeB-Images Modeling Oersted Fields}.(d). Even for $I_p = 5$ mA the Oersted field is still about $50\%$ of the coercive field, and so is still likely to influence the switching process.

With recent interest in resolving the contributions of several mechanisms to the switching of similar devices, this result shows that Oersted fields may not always be discounted when interpreting the relative contribution of each switching mechanism, and also that care must be taken when designing devices of this type. A simple method of minimising $B_y$, so as to explore only the spin-torque contributions to the switching, is therefore to perform a post deposition etch to remove CoFeB at the edges of the Hall bar, leaving only the Ta underlayer. In the majority of recent studies the HM underlayer has been thicker than the FM layer and so carries the majority of current, meaning that perpendicular Oersted fields in the FM would be effectively minimised by this approach. Another approach to studying the spin transfer torque is to apply an in-plane external magnetic field to the structure, so that switching occurs via coherent rotation\cite{Hao2015} of magnetization rather than by domain nucleation and growth. 

\section{\label{sec:level1}Summary}

Current induced switching in perpendicularly magnetised Ta/CoFeB/MgO layers was studied by simultaneous Kerr microscopy and electrical transport measurements, focusing on currents close to the critical value for switching. For zero applied magnetic field we find the switching to be a stochastic domain wall driven process, the speed of which is strongly dependent upon the value of the applied current. The nucleation of reverse domains appears to begin at one edge of the device, before these domains then grow towards the center of the Hall bar. Modelling the Oersted field through the cross section of the Hall bar reveals that the out of plane component is comparable to the 10 Oe out-of-plane coercive field of the CoFeB, suggesting that the Oersted field may assist the initial domain nucleation on one edge of the Hall bar while opposing reversal on the other edge. 

With recent interest in utilising Ta/CoFeB/MgO layers in perpendicular magnetic tunnel junctions this study highlights the need for careful consideration of the Oersted field when analysing potential contributions to the switching process. Minimisation of the Oersted field contribution, to facilitate study of spin-torques, can be achieved by etching the CoFeB layer at the edge of the device, although it may be possible to also utilise these fields to improve switching efficiency in future technologies.\\ 

\section*{Acknowledgements}
The authors wish to thank Tom Loughran, Rob Valkass, Paul Keatley, Wenzhe Chen and Shu-tong Wang for assistance and discussion. This work was supported by the University of Exeter through the award of a scholarship to CJD, and at Brown University by the Nanoelectronics Research Initiative (NRI) through the Institute for Nanoelectronics Discovery and Exploration (INDEX), and by the National Science Foundation through Grant No. DMR-1307056.

\bibliographystyle{unsrt}
\bibliography{PhDThesis}

\end{document}